# No Trust without regulation!
*European challenge on regulation, liability and standards for trusted AI*

*François Terrier*

Institut LIST, CEA, Université Paris-Saclay, F-91120, Palaiseau, France
francois.terrier@cea.fr

The explosion in the performance of Machine Learning (ML) and the potential of its applications are strongly encouraging us to consider its use in industrial systems, including for critical functions such as decision-making in autonomous systems. While the AI community is well aware of the need to ensure the trustworthiness of AI-based applications, it is still leaving too much to one side the issue of safety and its corollary, regulation and standards, without which it is not possible to certify any level of safety, whether the systems are slightly or very critical.

The process of developing and qualifying safety-critical software and systems in regulated industries such as aerospace, nuclear power stations, railways or automotive industry has long been well rationalized and mastered. They use well-defined standards, regulatory frameworks and processes, as well as formal techniques to assess and demonstrate the quality and safety of the systems and software they develop. However, the low level of formalization of specifications and the uncertainties and opacity of machine learning-based components make it difficult to validate and verify them using most traditional critical systems engineering methods. This raises the question of qualification standards, and therefore of regulations adapted to AI. With the AI Act, the European Commission has laid the foundations for moving forward and building solid approaches to the integration of AI-based applications that are safe, trustworthy and respect European ethical values. The question then becomes *"How can we rise to the challenge of certification and propose methods and tools for trusted artificial intelligence?"*

Dr. François Terrier,

Director of the Programs of the CEA's Institute for smart digital system (CEA List)

*François Terrier is AI Senior Fellow at CEA. He has a PhD in artificial intelligence and worked 10 years in the domain of expert systems using three-valued, temporal or fuzzy logics. Since 1994, he directs research on system and software engineering. As head of the system and software engineering department of the CEA, he led actions to build, for trustworthy software and systems, open tool chains covering the whole development cycle from requirement specification until equipment integration. His research challenges are, namely, on combining domain oriented modeling with formal methods for high quality, safe and secure critical systems. In 2019, François has been in charge to build the Trustworthy AI program of CEA and becomes in 2022 the Director of the Programs of CEA List (Institute for smart digital systems).*

Invited talks: 3rd Tailor workshop, 2023-06-06 ; AI Safety workshop of IJCAI 2023, 2023-08-21

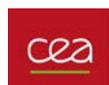

**Safety and security: a systems engineering perspective on trust**

The explosion in the performance of Machine Learning (ML) and the potential of its applications are strongly encouraging consideration of its use in industrial systems, including for critical functions such as decision-making in autonomous systems this leads an urgent need to consider safety as a key subject for any AI based system.

Establishing trust in AI-based systems is a strategic objective set by the European Commission for the massive deployment of AI [2]. In this vision, trust is linked to a set of different characteristics such as: data quality, fairness, traceability, explainability/interpretability, robustness, information protection, uncertainty or accuracy, etc. The natures and objectives of these items are not yet perfectly defined, and sometimes overlap or even conflict.

However, when we look at the development of AI-based systems involved in potentially critical functions or uses, we can distinguish between two points of view:

- That of the acceptability of the functions and uses of these systems and
- That of the certification/qualification of systems for a given field of use.

These two points of view complement each other, but are based on communities with cultures that are still very different and not very integrated: that of systems engineering, in particular for safety engineering and qualification, and that of the development, essentially software, of AI applications.

- While the AI community is well aware of the need to ensure the trustworthiness of AI-based applications, it is still leaving too much to one side the issue of safety and its corollary, regulation and standards, without which it is not possible to certify any level of safety, whether the systems are slightly or very critical.
- The process of developing and qualifying safety-critical software and systems in regulated industries such as aerospace, nuclear power stations, railways and even the automotive industry has long been well rationalised and mastered. These industries use well-defined standards, regulatory frameworks and processes, as well as formal techniques to assess and demonstrate the quality and safety of the systems and software they develop.

Considering these two viewpoints (usage and engineering) make it possible to classify the characteristics according to the main issues and priority points of attention of these two communities. Typically,

- On the usage side, a strong focus is placed on acceptability, with a preponderance of interest in characteristics directly related to the human end-user, such as explainability (to the user), transparency, fairness and, very quickly, questions of societal impact and ethics.
- On the engineering side, the focus will be more on the ability to qualify and certify the system in terms of operating safety, reliability, cybersecurity and data protection, robustness, uncertainty, traceability and interpretability (in the sense of prediction, understanding of internal behaviour), in particular.

They are of course interdependent and the appropriate treatment of each of them is necessary for the production of trustworthy systems. However, when it comes to integrating AI into mission-critical systems, the engineering viewpoint predominates when it comes to setting up qualification and certification processes.

**No Trust without regulation!** *European challenge on regulation, liability and standards for trusted AI*

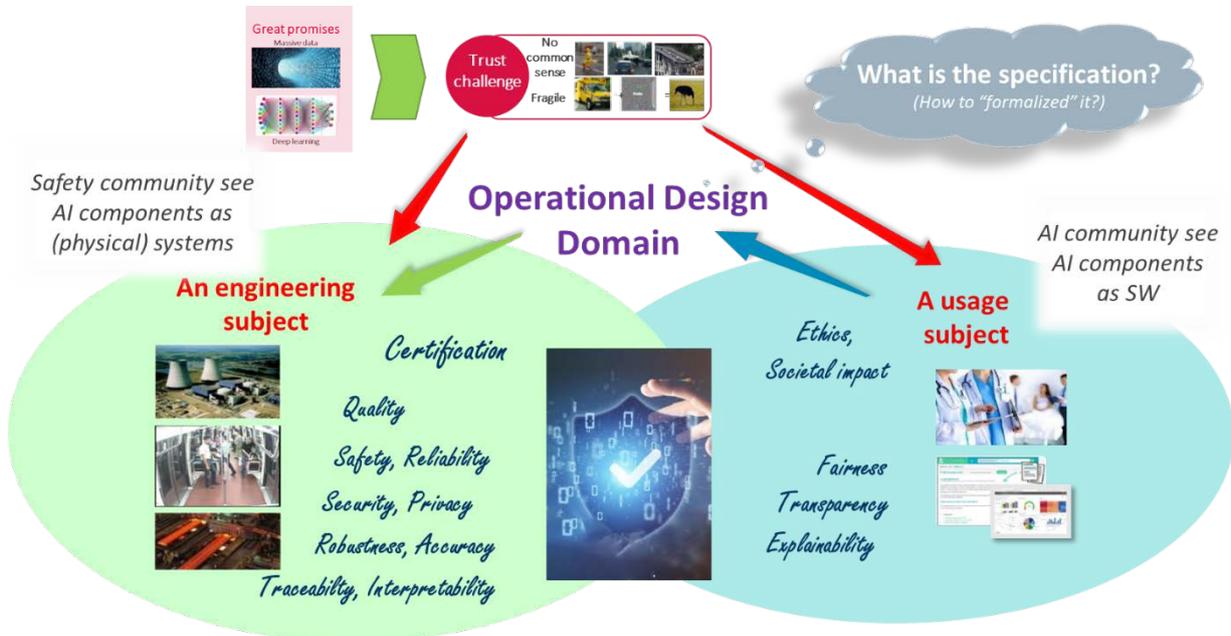

*TRUST a potential showstopper for AI DEPLOYMENT*

One of the great strengths and breakthroughs of AI is its ability to generate solutions to complex problems on the basis of a collection of more or less structured information, of which learning-based methods are outstanding examples. By its very nature, the development of AI-based systems relies heavily on highly empirical approaches to collecting information (data, expert rules, knowledge, etc.) and then building the application, often in a highly experimental manner. This flexibility is also its greatest weakness compared with the usual engineering practices for mission-critical systems, which are based on formalising specifications and applying them throughout development in a traceable and verifiable way right through to final implementation.

So, for example, any value placed in a conventional programme can be justified in terms of the initial requirements that produced it. Conversely, tracing the reason for the final value of a hyper-parameter in a neural network in relation to an initial requirement will generally be impossible both because of the lack of exploitable traceability and because of the weakness of the formalisation of the expected function (or 'task'). Similar difficulties will be encountered in systems based on symbolic AI, at the very least with regard to formalising the specification of the function, due to the absence of successive refinement of the requirements leading to the elementary knowledge elements assembled.

However, the low level of formalisation of specifications and the uncertainties and opacity of machine learning-based components make it difficult to validate and verify them using most traditional methods of critical systems engineering. This raises the question of qualification standards, and therefore of regulations adapted to AI. With the AI Act, the European Commission has laid the foundations for moving forward and building solid approaches to the integration of AI-based applications that are safe, trustworthy and respect European ethical values. The question then becomes : *How can we rise to the challenge of certification and propose methods and tools for trusted artificial intelligence?*

It was to address this tension between the requirements of established safety engineering practices and AI development practices that the CEA initiated a programme strategy on trusted AI and its industrial deployment in 2018 [1]. This has motivated the development of communities bringing together AI and safety engineering players through various workshops, theme days [4] and European networks of excellence such as Tailor [5]. It should be noted that these initiatives immediately found large and lasting audiences, since they have been going on for 4 years, underlining the expectations and the importance of the subject.

**No Trust without regulation!** *European challenge on regulation, liability and standards for trusted AI*

Following on from the exchange initiatives, a key issue in this field is the link with standards. This fundamental link is a particular focus of attention for the European Commission, which wants to equip itself as quickly as possible with the standards that will provide the operational means to implement its regulations [6]. Therefore, in addition to its participation in ISO and CEN/CENELEC committees, the CEA is involved in the European Adra-e project [7], in close coordination with the European Commission and its reference body CEN/CENELEC, to monitor and recommend IA standards and regulations.

In practical terms, the implementation of qualification solutions involves more than simply adapting or using off-the-shelf bricks. It requires in-depth methodological reflection and the identification of solutions appropriate to the various activities, the AI technologies used and the challenges of use. This is what motivated the French government to launch Confiance.ai [8], a major national programme on trust in critical systems, which is particularly instrumental in building a common understanding and approach to the broad and secure integration of AI in industrial systems.

In this context, the approach adopted at the CEA is to pursue broad-based R&D in order to identify promising elements and to mobilise the skills of available researchers through targeted and complementary actions. However, there are a number of key areas of focus, based on pre-existing expertise and anchored in teams already established in the fields of safety engineering and artificial intelligence.

*Risk analysis and operational design domain*

The qualification approach requires a definition of the risks and specification of the system's operating context. This applies to AI-based systems, with the need to integrate the specific elements of these systems into the analysis. With this objective in mind, work is being carried out to support the qualification process by creating a repository, a model, of the risks to be integrated into the safety analysis and design of AI-based systems [9]. This analysis contributes to the specification of the operational design domain (ODD), which is a key to the formalisation of AI-based system specifications [10].

For the analysis itself, many factors come into play, such as the qualification of data, the design of AI that is robust by construction, and the assessment of uncertainty in models. A key stage for critical systems is the verification and validation of AI-based systems. Test techniques are still fundamental at this stage of the state of the art, but they are unsatisfactory and very difficult to use because of the size of the state space to be covered and the difficulty of defining notions of coverage of this space. To compensate, in part, for these test validation difficulties, the usual critical systems make extensive use of formal methods. The question is whether these approaches can be applied to the technical domain of AI.

*Formal methods and AI: two worlds that can be reconciled*

For many years now, formal methods have been very well integrated into the development processes for critical systems and the associated certification procedures. They make it possible to move on from statistical, or even empirical, approaches to verification by testing, by providing "proof" and coverage arguments for the validation of software systems. However, their application to the field of AI raises questions about their ability to be scaled up to the AI applications they are intended for, and about their real areas of use: what are we trying to validate? at what stage of development? in relation to what characteristics of trust?

Formal methods thus find themselves back in the situation of the 1980s, when the computer development community did not see their applicability in the short term, or even, as some very high-level researchers said, that they would never be operational on real programs: "program verification is bound to fail. We can't see how it's going to be able to affect anyone's confidence about programs" [11]. Since then, the situation has evolved to such an extent that they are recommended in certification processes in various critical systems domains (aeronautics, railways, nuclear, etc.) [12].

**No Trust without regulation!** *European challenge on regulation, liability and standards for trusted AI*

The fundamental question remains the definition of the properties to be verified and the contribution of these verifications to the overall process. To help frame the use of these approaches, three situations can be distinguished:

- *Applications based on symbolic AI*, which by their very nature are constructed from a set of elements (rules, constraints, etc.) that can be semantically and mathematically interpreted without ambiguity. The difficulty lies in the sheer number of elements associated in a single application, often without any hierarchical or compositional structure. They also often suffer from a lack of specification of the global functions targeted and system properties to be verified. However, these points can be addressed in a joint development process from the outset between the AI development teams and validation teams familiar with formal methods. This can go as far as supporting development teams in the use of formal design languages that provide, by construction, proofs of consistency, completeness or safety properties. This is an approach developed by the CEA, notably through the Colibri tool [13].

- *Formally specified learning-based applications* are typically control-command applications, in the sense of automation, an illustrative case of which in the AI community is the anti-collision trajectory control system for aircraft, known as ACAS-XU . In this context, we benefit from a mathematical definition of the specifications and properties to be verified, which can be exploited by formal methods, and the question is to know whether these specifications are satisfied by the learning-based implementation. Initial work shows that this validation is theoretically achievable, but that it comes up against problems of combinatorics, classic of formal methods, which require work on optimising the solvers and associating the building blocks with the best state-of-the-art performance.

    An interesting case is the use of simulation for the development of a learning-based system. In this context, the simulator itself can constitute the formal specification of the system and opens up the possibility of formal demonstration of the learning-based system with respect to this specification [14].

- *Learning-based applications without formal specification*. In this case, formal analysis can also be used for robustness analysis, in the sense of verifying the stability of outputs in the face of input perturbations. In this context, the formalisation concerns the definition of these inputs (e.g. intervals of values, geometric properties between inputs, or typical disturbance profiles such as luminosity). The analysis then seeks to prove that the decisions taken by the AI-based system are, or are not, insensitive to these disturbances. Recent work has also demonstrated the ability to construct elements of understanding/interpretation/explanation of the operation of neural networks through abstract interpretation by automatically identifying key characteristics of the inputs used for classification [15].

    *These two approaches (verification of properties, robustness analysis) were implemented on an industrial use case of anchor line rupture control on Technip Energie's offshore drilling platforms. The results formally demonstrated its safe operating domain (compliance with its specification) and its robustness in the face of disturbances to input data that are realistic in terms of the environment.*

With all of this in mind, the CEA is developing the CAISAR integrative formal verification platform in order to facilitate the use of all the verification components at the highest level of the state of the art and in line with the targeted validation objective [16].

***The case of embedded systems: starting from scratch!***

Deploying AI in embedded systems is a pressing objective, and a particularly important one for critical systems that are often based on embedded functions. Obviously, all the above

**No Trust without regulation!** *European challenge on regulation, liability and standards for trusted AI*

issues apply here, with the addition of the need to deal with an extra stage after the functional design of the application: that of its coding in embedded systems. This last phase leads to a set of additional questions to be addressed:

- Coding optimisation in terms of resources: how can we check the impact of this optimisation on the previous checks? Do we know how to check that the optimised form conforms to the intermediate forms?
- Code deployment: what is the impact of dedicated AI calculation libraries, parallelization, dedicated hardware functions, calculation control (particularly parallelization) by a system operating system and cohabitation with other functions (programmes) of varying nature and criticality?
- Hardware faults: a classic embedded issue, we need to analyse the impact of these faults on AI applications, particularly those that require a lot of memory resources. What error detection and correction strategies should be adopted?
-

**Prospects and opportunities for breakthroughs**

Engineering the dependability of AI-based systems is still in its infancy, but by building on the maturity it has acquired in the field of software-based systems, it seems possible to adapt it fairly quickly.

One subject that is still largely prospective is that of distributed AI (and in particular decentralised learning), which still requires a detailed understanding of the issues, contributions (beyond the mere localisation of data close to uses and data) and impact of distribution in terms of the qualifications of AI-based systems.

In the longer term, the challenge remains to gain better control over the design of AI-based systems, and in particular to gain a detailed understanding of the learning mechanisms that enable behaviour to be predicted, and to obtain approaches that are more economical in terms of data, computation and resources (energy, materials, etc.) in order to implement frugal AI approaches.